\documentclass[aps,preprint]{revtex4}%
\usepackage{amsfonts}
\usepackage{amsmath}
\usepackage{amssymb}
\usepackage{graphicx}%
\providecommand{\U}[1]{\protect\rule{.1in}{.1in}}

\begin{document}
\title[ ]{Derivations of the Planck Blackbody Spectrum from Thermodynamic Ideas in
Classical Physics with Classical Zero-Point Radiation}
\author{Timothy H. Boyer}
\affiliation{Department of Physics, City College of the City University of New York, New
York, New York 10031}
\keywords{}
\pacs{}

\begin{abstract}
Based upon thermodynamic ideas, two new derivations of the Planck blackbody
spectrum are given within classical physics which includes classical
zero-point radiation. \ The first and second laws of thermodynamics, applied
to a harmonic oscillator or a radiation normal mode, require that the
canonical potential $\phi(\omega/T)$ is a function of a single variable
corresponding to the ratio of the oscillation frequency to the temperature.
\ The second law of thermodynamics involves extremum ideas which may be
applied to thermal radiation. \ Our first derivation of the Planck spectrum is
based upon the idea that the canonical potential $\phi(\omega/T)$ is a
monotonic function and all its derivatives are monotonic when interpolating
between zero-point energy at low temperature and energy equipartition at high
temperature; the monotonic behavior precludes the canonical potential from
giving a preferred value for the ratio $\omega/T.$ \ Our second derivation of
the Planck spectrum is based upon the requirement that the change in the
Helmholtz free energy of the radiation in a partitioned box held at constant
temperature should be a minimum at thermal equilibrium. \ Finally, the change
in Casimir energy with change in partition position for the radiation in a
partitioned box is shown to correspond at high temperature to the absence of
zero-point energy when the spectral energy per normal mode is chosen as the
traditional Planck spectrum which omits zero-point energy at low temperature;
thus the idea of zero-point energy is embedded in the traditional Planck
spectrum. \ \ It is emphasized that thermal radiation is intimately connected
with zero-point radiation and the structure of spacetime in classical physics.

\end{abstract}
\maketitle

\section{Introduction}

\subsection{False Claims in the Physics Literature}

The physics literature claims that attempts to explain the blackbody spectrum
within classical physics illustrate the breakdown of classical
physics.\cite{texts} \ However, this claim that the blackbody spectrum cannot
be explained within classical physics is simply erroneous. \ There have been a
number of valid derivations of the blackbody radiation spectrum within
classical physics.\cite{without}\cite{fluct}\cite{dia}\cite{acc} \ In the
present article, we present two new derivations from thermodynamic points of view.

The claims in the physics literature that classical physics cannot explain the
blackbody spectrum are a century out of date because they fail to consider the
two crucial aspects needed for understanding the phenomenon. \ These missing
aspects include: 1) the presence of classical electromagnetic zero-point
radiation, and 2) the importance of special relativity. \ The experimentally
observed Casimir forces\cite{Casimir}\cite{Cf} between conducting parallel
plates indicate unambiguously the presence of classical electromagnetic
zero-point radiation with a Lorentz-invariant spectrum. \ Of course, those
physicists who prefer to discuss physics within the context of
\textit{quantum} theory will describe the Casimir forces in terms of
\textit{quantum} zero-point radiation. \ However, if one is working within
\textit{classical} theory, then the presence of Casimir forces requires the
presence of \textit{classical} electromagnetic zero-point radiation.\cite{any}
\ In order to fit the experimental data on Casimir forces, the spectrum of
classical electromagnetic zero-point radiation must be Lorentz-invariant,
scale invariant, and indeed conformal invariant.\cite{Min} \ The one free
parameter regarding classical zero-point radiation is the multiplicative scale
factor which is chosen to fit the experimental data; the scale factor gives an
energy per normal mode of $(1/2)\hbar\omega$ where $\omega$ is the angular
frequency of the mode and $\hbar$ is a numerical constant which takes the same
value as Planck's constant.

\subsection{The Influence of Classical Zero-Point Radiation}

The presence of classical electromagnetic zero-point radiation will influence
all phenomena to a greater or lesser extent. \ Since there is classical
electromagnetic zero-point radiation present in the universe according to
classical theory, then classical statistical mechanics (with its assumption
that all motion stops at the absolute zero of temperature) is no longer valid
because zero-point radiation drives all electromagnetic systems into random
oscillation; rather, classical statistical mechanics can be regarded as simply
a large-mass-low-velocity approximation to thermal behavior where the
influence of classical zero-point radiation is small. \ Accordingly, and
contrary to what is claimed in the physics literature, the Rayleigh-Jeans law
for thermal radiation is not the unique result of classical theory, but is
merely the spectrum holding at long wavelength and low temperatures where
zero-point radiation has small influence. \ 

Starting from the presence of classical electromagnetic zero-point radiation,
several derivations of Planck's spectrum for blackbody radiation have been
given. \ These include discussion of the motion of a dipole oscillator in a
box (analogous to the discussion of Einstein and Stern),\cite{without} the
treatment of thermal fluctuations above the zero-point
fluctuations,\cite{fluct} the use of free-particle diamagnetism in the
large-mass-low-velocity limit when classical zero-point radiation is
present,\cite{dia} and use of a time-dilating conformal transformation of
classical zero-point radiation in a Rindler frame.\cite{acc}

\subsection{Two New Derivations of the Blackbody Spectrum within Classical
Physics}

In the present article, we offer two new derivations of the blackbody
radiation spectrum within classical physics based upon thermodynamic ideas.
\ Both derivations depend upon the presence of classical electromagnetic
zero-point radiation. \ 

The first part of the thermodynamic analysis is used subsequently in both
derivations. \ We start by applying the first two laws of thermodynamics to a
harmonic oscillator system and find that all the thermodynamic functions for
the oscillator depend upon one unknown canonical potential
function\cite{Garrod} $\phi(\omega/T)$ depending upon the single variable
$\omega/T$ corresponding to the ratio of the harmonic oscillator frequency
$\omega$\ to the temperature $T$. \ The energy $U(\omega,T)=-\omega
\phi^{\prime}(\omega/T)$\ of the oscillator has zero-point energy and energy
equipartition as its asymptotic limits, and the full thermal behavior
corresponds to the interpolating function between these two limits. \ The
interpolating function must be determined in connection with the extremum
ideas of the second law of thermodynamics. \ 

Our first derivation of the Planck spectrum for a harmonic oscillator is based
upon the assumption of \textquotedblleft thermodynamic
smoothness,\textquotedblright\ that the canonical potential function
$\phi(\omega/T)$ is monotonic and that all its derivatives are monotonic, so
as to remove the possibility of a preferred value for \ $\omega/T$. \ The
assumption about monotonic behavior sharply restricts the class of functions
allowed in the interpolation, and it is possible to pick out the interpolation
function from the restricted class of functions. \ The Planck spectrum indeed
satisfies the required condition.

Our second derivation applies the minimum principle for the Helmholtz free
energy to the thermal scalar radiation trapped in a one-dimensional box with a
partition. If we choose a test interpolating function $\phi_{t}(\omega/T)$ for
a single radiation mode, then we can calculate the functional dependence for
the change in the Helmholtz free energy (involving infinitely many modes in
the box) for fixed temperature as the position of the partition is altered.
\ We see that most interpolation functions $\phi_{t}(\omega/T)$ for a mode do
not satisfy the minimum principle for the Helmholtz free energy for the
partition at the center of the box for a box of arbitrary length.
\ However,\ we find that indeed the Planck spectrum satisfies the required
minimum principle for all box lengths.

Finally, we calculate the change in Casimir energy for a partitioned box. \ We
show that the traditional Planck spectrum which omits zero-point radiation at
low temperature does not go over at high temperature to the expected
Rayleigh-Jeans result. \ On the other hand, the full Planck spectrum which
includes zero-point radiation at low temperature does indeed go to the
Rayleigh-Jeans result at high temperature.

\section{Second Law of Thermodynamics and Zero-Point Energy}

\subsection{The First and Second Laws of Thermodynamics Applied to the
Harmonic Oscillator}

We start by considering the thermodynamics of the harmonic oscillator, since a
small harmonic oscillator comes to equilibrium with thermal radiation at the
same average energy as the radiation normal mode at the same frequency as the
oscillator.\cite{Lavenda} \ Alternatively, we can think of a radiation mode as
behaving like a harmonic oscillator.

Now the thermodynamics of a harmonic oscillator has only two thermodynamic
variables $T$ and $\omega,$ and takes a particularly simple form.\cite{SHO}
\ In thermal equilibrium with a bath, the average oscillator energy
$\left\langle \mathcal{E}\right\rangle $ is denoted by $U=\left\langle
\mathcal{E}\right\rangle =\left\langle J\right\rangle \omega,$ and satisfies
$dQ=dU+dW$ with the entropy $S$ satisfying $dS=dQ/T.$ \ Since $J$ is an
adiabatic invariant for the oscillator,\cite{Gold} the work done by the system
is given by $dW=-\left\langle J\right\rangle d\omega=-(U/\omega)d\omega.$
\ Combing these equations, we have $dS=dQ/T=[dU-(U/\omega)d\omega]/T.$
$\ $Writing the differentials in terms of $T$ and $\omega,$ we have
$dS=(\partial S/\partial T)_{\omega}dT+(\partial S/\partial\omega)_{T}d\omega$
and $dU=(\partial U/\partial T)_{\omega}dT+(\partial U/\partial\omega
)_{T}d\omega.$ \ Therefore ($\partial S/\partial T)_{\omega}=(\partial
U/\partial T)_{\omega}/T$ and ($\partial S/\partial\omega)_{T}=[(\partial
U/\partial\omega)_{T}-(U/\omega)]/T.$ \ Now equating the mixed second partial
derivatives $\partial^{2}S/\partial T\partial\omega=\partial^{2}%
S/\partial\omega\partial T,$ we have $(\partial^{2}U/\partial\omega\partial
T)/T=(\partial^{2}U/\partial T\partial\omega)/T-(\partial U/\partial
T)_{\omega}/(T\omega)+[(U/\omega)-(\partial U/\partial\omega)_{T}]/T^{2}$ or
$0=-(\partial U/\partial T)_{\omega}/(T\omega)+[(U/\omega)-(\partial
U/\partial\omega)_{T}]/T^{2}.$ \ The general solution of this equation is
$U=\omega f(\omega/T)=\omega\left\langle J\right\rangle ~$where $f\left(
\omega/T\right)  $ is an unknown function which corresponds to the average
value $\left\langle J\right\rangle $ of the action variable of the oscillator.
\ If we had equated the mixed partial derivatives of the energy, then we find
the equation ($\partial S/\partial\omega)_{T}=-(T/\omega)(\partial S/\partial
T)_{\omega},$\ which has the general solution $S(\omega,T)=g(\omega/T)$ where
$g$ is an arbitrary function. \ The information provided by the second law of
thermodynamics is that there is a single function $\phi(\omega/T)$
corresponding to the canonical potential function\cite{Garrod} which gives the
Helmholtz free energy as
\begin{equation}
F(\omega,T)=-T\phi(\omega/T), \label{F}%
\end{equation}
the average oscillator energy as%
\begin{equation}
U(\omega,T)=T^{2}\left(  \frac{\partial\phi}{\partial T}\right)  _{\omega
}=-\omega\phi^{\prime}(\omega/T), \label{U}%
\end{equation}
and the entropy as%
\begin{equation}
S(\omega/T)=\phi(\omega/T)+U(\omega,T)/T=\phi(\omega/T)-(\omega/T)\phi
^{\prime}(\omega/T). \label{S}%
\end{equation}
Thus the thermodynamics of the harmonic oscillator is determined by one
unknown function $\phi(\omega/T).$ \ When applied to thermal radiation, the
result obtained here purely from the first and second laws of thermodynamics
corresponds to the familiar Wien displacement law of classical physics.

\subsection{Possibility of Zero-Point Energy and Zero-Point Radiation}

The energy expression (\ref{U}) for a harmonic oscillator (or an
electromagnetic radiation mode) in thermal equilibrium allows two limits which
make the energy independent from one of its two thermodynamic variables.
\ When the temperature $T$ becomes very large so that the ratio $\left(
\omega/T\right)  $ is small, the average energy $U$\ of the mode in Eq.
(\ref{U}) becomes independent of the frequency $\omega$ provided $\phi
^{\prime}(\omega/T)\rightarrow-const_{1}\times(\omega/T)^{-1}$ so that%

\begin{equation}
U=-\omega\phi^{\prime}(\omega/T)\rightarrow-\omega\times\lbrack-const_{1}%
\times T/\omega]=const_{1}\times T\text{ \ \ for \ }\omega/T<<1. \label{Uh}%
\end{equation}
This is the familiar high-temperature limit where we expect to recover the
Rayleigh-Jeans equipartition limit. \ Therefore we choose this constant as
$const_{1}=k_{B}$ corresponding to Boltzmann's constant. \ With this choice,
our thermal radiation now goes over to the Rayleigh-Jeans limit for high
temperature or low frequency.

In the other limit of small temperature where the ratio $\omega/T$ is large,
the dependence on temperature is eliminated provided $\phi^{\prime}%
(\omega/T)\rightarrow-const_{2},$ so that
\begin{equation}
U=-\omega\,\phi^{\prime}(\omega/T)\rightarrow-\omega\times\lbrack
-const_{2}]=const_{2}\times\omega\text{ \ \ for \ }\omega/T>>1. \label{Ul}%
\end{equation}
At this point, any theoretical description of thermal radiation involves a
choice, which should be based on experimental observation. \ If we choose this
second constant to vanish, $const_{2}=0,$ then this limit does not force us to
introduce any constant beyond Boltzmann's constant, which entered for the
high-temperature limit of thermal radiation. \ On the other hand, if we choose
a non-zero value for this constant, $const_{2}\neq0,$ then we are introducing
a second constant into the theory of thermal radiation, which constant has
different dimensions from those of Boltzmann's constant. \ The units of this
new constant $const_{2}$ correspond to \textit{energy} times \textit{time}.
\ Furthermore, the choice of a non-zero value for this constant means that at
temperature $T=0,$ there is random, temperature-independent energy present in
the harmonic oscillator. \ If this harmonic oscillator has electromagnetic
interactions, it must be in equilibrium with the radiation in the thermal
bath, and therefore random zero-point radiation must be present in the system.
\ This random radiation which exists at temperature $T=0$ is classical
electromagnetic zero-point radiation. \ 

We emphasize that thermodynamics allows classical zero-point radiation within
classical physics. \ The physicists of the early 20th century were not
familiar with the idea of classical zero-point radiation, and so they made the
assumption $const_{2}=0$ which excluded the possibility of classical
zero-point radiation. \ In his monograph on classical electron theory,
Lorentz\cite{Lorentz} makes the explicit assumption that there is no radiation
present at $T=0.$ \ Today, we know that the exclusion of classical zero-point
radiation is a poor choice. \ However, the current textbooks of modern physics
continue to present only the outdated, century-old view.\cite{texts}

Once the possibility of classical zero-point radiation is introduced into
classical theory, one looks for other phenomena where the zero-point radiation
will play a crucial role. \ In particular, the Casimir force\cite{Casimir}
between two uncharged conducting parallel plates will be influenced by the
presence of classical electromagnetic zero-point radiation. \ By comparing
theoretical calculations with experiments, one finds that the scale constant
for classical zero-point radiation appearing in Eq. (\ref{Ul}) must take the
value $const_{2}=1.05\times10^{-34}$Joule-sec. \ However, this value
corresponds to the value of a familiar constant in physics; it corresponds to
the value $\hbar/2$ where $\hbar$ is Planck's constant. \ Thus in order to
account for the experimentally observed Casimir forces between parallel
plates, the scale of classical zero-point radiation must be such that
$const_{2}=\hbar/2,$ and for each normal mode, the average energy becomes
\begin{equation}
U=-\omega\phi^{\prime}(\omega/T)\rightarrow(\hbar/2)\omega\text{ \ \ for
}T\rightarrow0. \label{zpr}%
\end{equation}

We emphasize that Planck's constant enters classical electromagnetic theory as
the scale factor in classical electromagnetic zero-point radiation. \ There is
no connection whatsoever to any idea of quanta. \ Many physicists are misled
by the textbooks of modern physics and regard Planck's constant as a
\textquotedblleft quantum constant.\textquotedblright\cite{essay} \ This is a
completely misleading idea. \ A physical constant is a numerical value
associated with certain aspects of nature; the constant may appear in several
different theories, just as Cavendish's constant G appears in both Newtonian
physics and also in general relativity. \ Indeed, Planck's constant
$h=2\pi\hbar$ was introduced into physics in 1899 before the advent of quantum
theory.\cite{DSB} \ Planck's constant can appear in both classical and quantum theories.

\section{Derivation of the Planck Spectrum Based upon the Idea of
Thermodynamic Smoothness}

\subsection{Choosing Constants Such that k$_{B}=1$ and $\hbar=1$}

When dealing with the thermodynamics of the harmonic oscillator, it is
convenient to absorb Boltzmann's constant $k_{B}$ into the definition of
temperature and to absorb Planck's constant into the definition of
frequency.\cite{units} \ In this convention, the two constants become
$const_{1}=1$ and $const_{2}=1/2.$ \ In the thermodynamic review above, we see
that the thermodynamics of the harmonic oscillator, and therefore of the
blackbody radiation spectrum is determined by one unknown function $\phi(z)$
where $z=\omega/T$ which has the asymptotic limits for its derivative given by%
\begin{equation}
\phi^{\prime}(z)\rightarrow-z^{-1}\text{ for }z\rightarrow0\text{ \ and
\ }\phi^{\prime}(z)\rightarrow-1/2\text{ for }z\rightarrow\infty. \label{fp}%
\end{equation}
The function $\phi$ itself then has the asymptotic limits determined by
integrating once, giving for $-\phi(z)$%
\begin{equation}
-\phi(z)\rightarrow\ln z\text{ for }z\rightarrow0\text{ \ and \ }%
-\phi(z)\rightarrow z/2\text{ for }z\rightarrow\infty\label{mp}%
\end{equation}
plus possible constants. \ 

\subsection{Thermodynamic Smoothness Applied to the Harmonic Oscillator}

In obtaining the results in Eqs. (\ref{fp}) and (\ref{mp}), we have used the
first and second laws of thermodynamics including the idea of an entropy
function $S$ which is a state function. However, the analysis does not include
the concept that the entropy function assumes a maximum value associated with
stability. \ This stability idea includes the notion of thermodynamic
smoothness which demands that the canonical potential for the oscillator does
not distinguish any frequency $\omega$ at a given temperature $T,$ nor any
temperature at a given frequency. \ At a minimum, the notion of smoothness
demands that any interpolation function $\phi(\omega/T)=\phi(z)$ for the
canonical potential of the oscillator is monotonic and all its derivatives are
monotonic; the monotonic behavior prevents a single value for $\omega/T$ from
being distinguished by the canonical potential.

Now the set of functions which are monotonic and all of whose derivatives are
monotonic is extremely limited. \ The set includes $x,$ $e^{x},~\sinh x,~\cosh
x,$ $\tanh x,$ their inverses and powers. \ In particular, we notice that the
hyperbolic sine function has the asymptotic limits%
\begin{equation}
2\sinh(z/2)\rightarrow z\text{ for }z\rightarrow0\text{ while }2\sinh
(z/2)\rightarrow e^{z/2}~\text{\ for }z\rightarrow\infty\label{sinh}%
\end{equation}
But this looks like exactly the exponentiation of the interpolation limits in
Eq. (\ref{mp}) which we required for the canonical potential function
$\phi(z).$ \ This suggests that the needed smooth interpolation function
$\phi_{Pzp}$ is given by
\begin{equation}
\phi_{Pzp}(z)=-\ln[2\sinh(z/2)] \label{phi}%
\end{equation}

We can check that $\phi_{Pzp}(z)$ given in Eq. (\ref{phi}) is indeed monotonic
and all its derivatives are monotonic. \ The function $\phi_{Pzp}(z)$ is
clearly monotonic since both $\ln x$ and $\sinh x$ are monotonic for $0<x$.
The first derivative of $\phi_{Pzp}(z)$ is%

\begin{equation}
\phi_{Pzp}^{\prime}(z)=-(1/2)\coth(z/2) \label{fpc}%
\end{equation}
This function also is monotonic, and, since $\coth x=1/x+x/3-x^{3}%
/45+2x^{5}/495-...\rightarrow1/x$ for $x\rightarrow0,$ while $\coth
x\rightarrow1,$ for $x\rightarrow\infty;$ thus we see that $\phi_{Pzp}%
^{\prime}(x)$ in Eq. (\ref{fpc}) has asymptotic limits in agreement with Eq.
(\ref{fp}). \ \ 

It is clear that we need to prove that all the derivatives of $\coth x$ are
monotonic. \ One method of proof uses an exponential expansion,%

\begin{align}
-\phi_{Pzp}^{\prime}(z)  &  =\frac{1}{2}\coth\left(  \frac{z}{2}\right)
=\frac{1}{2}+\frac{1}{\exp(z)-1}=\frac{1}{2}+\exp(-z)\frac{1}{1-\exp
(-z)}\nonumber\\
&  =1/2+e^{-z}+e^{-2z}+e^{-3z}+...
\end{align}
which involves a constant function and then a sum of functions all of which
are monotonically decreasing, so that the function is monotonically decreasing
in $z.$ \ By the ratio test, the series is absolutely convergent since
$0<e^{-z}<1$ for $0<z$. \ The second derivative gives%
\begin{equation}
-\phi_{Pzp}^{\prime\prime}(z)=-e^{-z}-2e^{-2z}-3e^{-3z}-
\end{equation}
and again all of the terms are of the same (negative) sign, and all are
monotonically decreasing in magnitude so the that function is monotonically
increasing. \ Again by the ratio test, the series is absolutely convergent
since $0<[(n+1)/n]e^{-z}<1$ for sufficiently large $n$ for fixed $z,$ $0<z$.
\ Indeed, it is easy to see that the pattern is repeated upon further
differentiation, so that the series expansion in terms of exponentials is
absolutely convergent by the ratio test and all derivatives of the canonical
potential $-\phi_{Pzp}(z)$ are monotonic.

\subsection{Smooth Interpolation Gives the Planck Function}

Now the canonical potential $\phi_{Pzp}$ for the harmonic oscillator which we
have obtained in Eq. (\ref{phi}) by assuming monotonic behavior between the
asymptotic limits is exactly that corresponding to the Planck formula
including zero-point energy. \ The Helmholtz free energy $F_{Pzp}$
corresponding to the Planck spectrum with zero-point energy for a harmonic
oscillator or a radiation mode of frequency $\omega$ is given by
\begin{equation}
F_{Pzp}(\omega,T)=-T\phi_{Pzp}(\omega/T)=T\ln\{2\sinh[\omega/(2T)]\}
\end{equation}
The associated energy $U_{Pzp}(\omega,T)$ follows as
\begin{equation}
U_{Pzp}(\omega,T)=T^{2}\left(  \frac{\partial\phi_{Pzp}}{\partial T}\right)
_{\omega}=-\omega\phi_{Pzp}^{\prime}(\omega/T)=\frac{\omega}{2}\coth\left(
\frac{\omega}{2T}\right)  =\frac{\omega}{2}+\frac{\omega}{\exp(\omega/T)-1}
\label{UPzp}%
\end{equation}
and the entropy $S_{Pzp}(\omega/T)$ as%
\begin{align}
S_{Pzp}(\omega/T)  &  =\phi_{pzp}(\omega/T)+U(\omega,T)/T=\phi_{Pzp}%
(\omega/T)-(\omega/T)\phi_{Pzp}^{\prime}(\omega/T)\nonumber\\
&  =-\ln\left[  2\sinh\left(  \frac{\omega}{2T}\right)  \right]  +\frac
{\omega}{2T}\coth\left(  \frac{\omega}{2T}\right)
\end{align}
The zero-point energy actually makes no contribution to the entropy $S_{Pzp}$
since in the limit of large $z,$ $\phi_{Pzp}(z)\rightarrow\phi_{zp}(z)=-z/2$
while%
\begin{equation}
S_{zp}(z)=\phi_{zp}(z)-z\phi_{zp}^{\prime}(z)=-z/2-z(-1/2)=0.
\end{equation}

\subsection{Entropy as a Monotonic Function of $U_{Pzp}/(\hbar\omega/2)$}

Since the first law of thermodynamics requires that the entropy $S(\omega
/T)$\ for a harmonic oscillator is a function of the single variable
$\omega/T,$ and also the energy of the oscillator is given by $U=-\omega
\phi^{\prime}(\omega/T)$ as in Eq. (\ref{U}), it follows (by using the inverse
function of $\phi^{\prime}(\omega/T))$ that the oscillator entropy $S$\ can
also be regarded as a function of the single variable $U/\omega$; \ The
variable $U/\omega$ runs from the constant value $U/\omega\rightarrow1/2$ when
$\omega>>T,$ to the value $U/\omega\rightarrow T/\omega$ when $T>>\omega
.~\ $We expect the oscillator entropy to be a monotonically increasing
function of temperature. \ Furthermore, the entropy should not distinguish any
preferred value of $U/\omega$. \ Thus we expect that the oscillator entropy
should be a monotonic function of $U/\omega$ and all its derivative should be
monotonic functions of $U/\omega.$ \ Indeed for the Planck relation given in
Eq. (\ref{phi}), we can use the fact that the inverse function for
$y=\coth(z/2)$ is\cite{AS87}
\begin{equation}
\frac{z}{2}=arc\coth(y)=\frac{1}{2}\ln\left(  \frac{y+1}{y-1}\right)  \text{
for }y^{2}>1
\end{equation}
where $y=U/(\omega/2)$ and $z=\omega/T$ to obtain
\begin{equation}
\frac{S_{Pzp}(y)}{\hbar k_{B}}=\frac{1}{2}(y+1)\ln(y+1)-\frac{1}{2}%
(y-1)\ln(y-1)-\ln2\text{ where }y=U_{Pzp}/(\hbar\omega/2) \label{SPzp}%
\end{equation}
By direct differentiation of the expression in Eq. (\ref{SPzp}), it is easy to
show that the Planck oscillator entropy $S_{Pzp}$ in Eq. (\ref{SPzp}) is
indeed a monotonic function of $U/\omega$ and all the derivatives are
monotonic functions.

This concludes our first derivation of the Planck blackbody spectrum based
upon thermodynamic ideas. \ For this derivation, we have discussed merely the
thermodynamics of a harmonic oscillator. \ For the second derivation of the
blackbody spectrum, we need to discuss radiation explicitly.

\section{ Relativistic Scalar Field Theory in One Spatial Dimension}

\subsection{Valid Thermodynamic Systems}

Blackbody radiation is the random radiation in an enclosure which is stable
under scattering. For a system involving radiation, there are infinitely many
normal modes of oscillation for the radiation so that the radiation must be
treated in the context of a relativistic field theory.

In order to understand blackbody radiation, we must choose models which do not
violate the principles of thermodynamics. \ There are several models mentioned
in the literature which obviously do violate the laws of thermodynamics. \ The
use of a Maxwell demon is the most famous example. \ However, the use of
nonrelativistic statistical mechanics with its energy equipartition for each
harmonic oscillator mode obviously violates the laws of thermodynamics when
applied to thermal radiation since it gives an ultraviolet divergence for the
energy. \ Similarly, using a nonrelativistic nonlinear dipole oscillator as a
radiation scatterer also violates the laws of thermodynamics when applied to
thermal radiation since the oscillator scatters the radiation toward the
equipartition result.\cite{scat} \ \ Although most physicists repeat the claim
that these thermodynamic failures arise due to the use of classical rather
than quantum physics, it has been suggested repeatedly and with ever more
convincing evidence that, insofar as classical physics is concerned, the
failure involves the invalid use of nonrelativistic physics together with a
relativistic radiation system.\cite{mixed} \ The naive combination of
nonrelativistic and relativistic physics leads to systems which violate the
laws of physics. \ Indeed, if we consider lifting a nonrelativistic harmonic
oscillator system within a relativistic accelerating Rindler frame, it is easy
to show that this mixture of nonrelativistic and relativistic physics violates
the laws of thermodynamics. \ In order to have a valid thermodynamic system
for relativistic radiation, we must insist that the interactions of the
radiation system do not violate any aspects of relativity. \ 

In our next derivation of the blackbody radiation spectrum, we will use ideas
which are usually associated with Casimir forces. \ We will consider the
thermodynamic system involving relativistic radiation in one spatial dimension
in a box which contains a partition. \ The radiation and boundary conditions
provide a fully relativistic system.

\subsection{Scalar Field Theory}

For simplicity of calculation, we will use relativistic scalar radiation in
one spatial dimension. \ The Lorentz-invariant spacetime interval is
$ds^{2}=g_{\mu\nu}dx^{\mu}dx^{\nu}$ with indices $\mu=0,1,$ and $x^{0}%
=ct,~x^{1}=x,$ so that $ds^{2}=c^{2}dt^{2}-dx^{2}$ \ The Lagrangian density
for the massless scalar field $\varphi$ is given by
\begin{equation}
\mathcal{L=}\frac{1}{2}\partial_{\mu}\varphi\partial^{\mu}\varphi=\frac{1}%
{2}\left[  \left(  \frac{\partial\varphi}{\partial ct}\right)  ^{2}-\left(
\frac{\partial\varphi}{\partial x}\right)  ^{2}\right]
\end{equation}
\ the stress-energy-momentum tensor density is%
\begin{equation}
T^{\mu\nu}=\partial^{\mu}\varphi\frac{\partial\mathcal{L}}{\partial
(\partial_{\nu}\varphi)}-g^{\mu\nu}\mathcal{L}%
\end{equation}
giving energy density%
\begin{equation}
u=T^{00}=\frac{1}{2}\left[  \frac{1}{c^{2}}\left(  \frac{\partial\varphi
}{\partial t}\right)  ^{2}+\left(  \frac{\partial\varphi}{\partial x}\right)
^{2}\right]
\end{equation}
and momentum density%
\begin{equation}
T^{01}=T^{10}=-\frac{1}{c}\frac{\partial\varphi}{\partial t}\frac
{\partial\varphi}{\partial x}%
\end{equation}

The equation of motion for the field corresponds to $\partial_{\mu}%
[\partial\mathcal{L}/\partial(\partial_{\mu}\varphi)]=0$%
\begin{equation}
\frac{1}{c^{2}}\frac{\partial^{2}\varphi}{\partial t^{2}}-\frac{\partial
^{2}\varphi}{\partial x^{2}}=0.
\end{equation}
If we choose to express the field $\varphi(ct,x)$ in a box running from $x=a$
to $x=b$ as a sum over normal modes which vanish at the ends (Dirichlet
boundary conditions), then
\begin{equation}
\varphi(ct,x)=\sum_{n=1}^{\infty}\varphi_{n}(ct,x)=\sum_{n=1}^{\infty}%
q_{n}(t)\left(  \frac{2}{b-a}\right)  ^{1/2}\sin\left[  \frac{n\pi}%
{b-a}(x-a)\right]  \label{nm}%
\end{equation}
where (using the orthogonality of the spatial normal mode functions) the
amplitude $q_{n}$ of the $n$th normal mode satisfies the differential
equation
\begin{equation}
\ddot{q}_{n}+\omega_{n}^{2}q_{n}=0 \label{SHO}%
\end{equation}
with
\begin{equation}
\omega_{n}=\frac{n\pi c}{b-a} \label{wDir}%
\end{equation}
This same frequency relation (\ref{wDir}) arises if we require that the first
spatial derivatives of the field vanish at the ends of the box (Neumann
boundary conditions), so that a cosine function replaces the sine function in
Eq. (\ref{nm}) . \ On the other hand, if we choose to express the field
$\varphi(ct,x)$ in terms of normal modes vanishing at $x=a$ (Dirichlet
boundary conditions) but with first spatial derivative vanishing at $x=b,$
(Neumann boundary conditions) then%
\begin{equation}
\varphi(ct,x)=\sum_{n=1}^{\infty}\varphi_{n}(ct,x)=\sum_{n=1}^{\infty}%
q_{n}(t)\left(  \frac{2}{b-a}\right)  ^{1/2}\sin\left[  \frac{(n-1/2)\pi}%
{b-a}(x-a)\right]
\end{equation}
where the amplitude $q_{n}$ again satisfies the harmonic oscillator
differential equation (\ref{SHO}), but the frequency is now
\begin{equation}
\omega_{n}=\frac{(n-1/2)\pi c}{b-a}. \label{wNeu}%
\end{equation}
When the boundary conditions (Dirichlet or Neumann) are the same at both ends
of the box, we speak of \textquotedblleft like boundary
conditions;\textquotedblright\ if the Dirichlet boundary conditions are used
at one end of the box and Neumann boundary conditions at the other, we speak
of \textquotedblleft unlike boundary conditions.\textquotedblright\ \ The
energy of the radiation in the box is given by%
\begin{align}
U  &  =\int_{x=a}^{x=b}dx\frac{1}{2}\left[  \frac{1}{c^{2}}\left(
\frac{\partial\varphi}{\partial t}\right)  ^{2}+\left(  \frac{\partial\varphi
}{\partial x}\right)  ^{2}\right] \nonumber\\
&  =\sum_{n=1}^{\infty}\mathcal{E}_{n}=\sum_{n=1}^{\infty}\frac{1}{2}(\dot
{q}_{n}^{2}+\omega_{n}^{2}q_{n}^{2}).
\end{align}
Thus each normal mode of the radiation field behaves like a harmonic
oscillator. \ 

\section{Derivation of the Planck Spectrum Based upon the Helmholtz Free
Energy for Radiation in a Partitioned Box}

\subsection{Thermal Radiation in a Box}

The simple harmonic oscillator equation of motion in (\ref{SHO}) can be solved
as $q_{n}(t)=f_{n}\cos(\omega_{n}t-\theta_{n})$ where $f_{n}$ gives the
amplitude of the oscillation and $\theta_{n}$ gives the phase. \ In the case
of thermal radiation in a box, the phases $\theta_{n}$ of the normal modes of
oscillation are completely uncorrelated so that we may write the radiation
field as%
\begin{equation}
\varphi(ct,x)=\sum_{n=1}^{\infty}\varphi_{n}(ct,x)=\sum_{n=1}^{\infty}%
f_{n}\left(  \frac{2}{b-a}\right)  ^{1/2}\sin\left[  \frac{\omega_{n}}%
{c}(x-a)\right]  \cos\left[  \omega_{n}t-\theta_{n}\right]
\end{equation}
where the phases $\theta_{n}$ are random variables which are independently
distributed for each normal mode $n.$ \ Thus when averaged in time or averaged
over the random phases, the averages involve%
\begin{equation}
\left\langle \sin(\omega_{n}t+\theta_{n})\sin(\omega_{n^{\prime}}%
t+\theta_{n^{\prime}})\right\rangle =\left\langle \cos(\omega_{n}t+\theta
_{n})\cos(\omega_{n^{\prime}}t+\theta_{n^{\prime}})\right\rangle
=(1/2)\delta_{nn^{\prime}}%
\end{equation}
while
\begin{equation}
\left\langle \sin(\omega_{n}t+\theta_{n})\cos(\omega_{n^{\prime}}%
t+\theta_{n^{\prime}})\right\rangle =0
\end{equation}
The amplitudes $f_{n}$ of the normal modes take values which are
characteristic of the frequency $\omega_{n}$\ of the mode and the temperature
$T$ of the box. \ Thus for thermal radiation, the average energy density
$u_{n}(x)$ of each normal mode $\varphi_{n}(cx,t)$ contributes separately to
the average energy density in the box, corresponding to
\begin{align}
\left\langle u_{n}(x)\right\rangle  &  =\frac{1}{2}\left\langle \frac{1}%
{c^{2}}\left(  \frac{\partial\varphi_{n}}{\partial t}\right)  ^{2}+\left(
\frac{\partial\varphi_{n}}{\partial x}\right)  ^{2}\right\rangle \nonumber\\
&  =\frac{1}{2}f_{n}^{2}\left(  \frac{\omega_{n}}{c}\right)  ^{2}\left(
\frac{2}{b-a}\right)  \frac{1}{2}\left\{  \sin^{2}\left[  \frac{\omega_{n}}%
{c}(x-a)\right]  +\cos^{2}\left[  \frac{\omega_{n}}{c}(x-a)\right]  \right\}
\nonumber\\
&  =\frac{1}{2}f_{n}^{2}\left(  \frac{\omega_{n}}{c}\right)  ^{2}\left(
\frac{2}{b-a}\right)  \frac{1}{2}%
\end{align}
Thus the average energy density is uniform across the box for each normal
mode, and the average total energy is given by
\begin{align}
U  &  =\sum_{n=1}^{\infty}\left\langle \mathcal{E}_{n}\right\rangle
=(b-a)\sum_{n=1}^{\infty}\frac{1}{2}f_{n}^{2}\left(  \frac{\omega_{n}}%
{c}\right)  ^{2}\left(  \frac{2}{b-a}\right)  \frac{1}{2}\nonumber\\
&  =\sum_{n=1}^{\infty}\frac{1}{2}f_{n}^{2}\left(  \frac{\omega_{n}}%
{c}\right)  ^{2}%
\end{align}

\subsection{Classical Zero-Point Radiation in a Partitioned Box}

The spectrum of zero-point energy corresponds to an average energy per normal
mode given by $U_{zp}(\omega)=(1/2)\hbar\omega.$ \ Since we are using units
where $\hbar=1,$ the zero-point energy simplifies to $U_{zp}(\omega
)=\omega/2.$ \ Classical electrodynamics, which involves a field theory in
three spatial dimensions, is invariant under Lorentz transformations and under
a single scale transformation characterized as $\sigma_{ltU^{-1}}$ where the
scale parameter $\sigma$ ranges over all positive values. \ The scale
transformation $\sigma_{ltU^{-1}}$ carries all lengths $l$ into $l^{\prime
}=\sigma l,$ all times $t$ into $t^{\prime}=\sigma t,$ and all energies $U$
into energies $U^{\prime}=U/\sigma.$ \ Such a scale transformation preserves
the values of the fundamental constants $c$ (the speed of light in vacuum)
having units of \textit{length/time}, $e$ (the charge of the electron) whose
square has units of \textit{energy} times \textit{length}, and $\hbar$ (the
scale factor for zero-point radiation) having units of \textit{energy} times
\textit{time}. \ The spectrum of zero-point radiation is Lorentz invariant and
also scale invariant.\cite{conf} \ It is easy to exhibit the scale invariance.
\ Under a scale transformation by a factor $\sigma,$ the relationship
$U_{zp}=\omega/2$ becomes $U_{zp}/\sigma=(\omega/\sigma)/2$ since the
frequency $\omega$ has units of 1/\textit{time}. \ But then we see that the
new relationship involves $U_{zp}^{\prime}=\omega^{\prime}/2\,\ $which is the
same as the original relationship.

The total zero-point energy of the radiation in a box clearly diverges since
there are infinitely many normal modes with ever-increasing frequency in the
box. \ However, as realized by Casimir in 1948, the change in zero-point
energy associated with a shift in the position of a partition in a box of
fixed length is indeed finite.\cite{Casimir} \ Thus we will consider a box of
total length $L$ containing a partition located at a distance $x$ from one
wall. \ The partition splits the original box of length $L$ into two boxes,
one of length $x$ and one of length $L-x.$ \ We consider the zero-point energy
$\mathcal{U}_{zp}(x)+\mathcal{U}_{zp}(L-x)$ in this partitioned box, and
compare it with the energy $2\mathcal{U}_{zp}(L/2)$ in the partitioned box
when the partition is half-way across the box of length $L$%
\begin{equation}
\Delta\mathcal{U}_{zp}=\mathcal{U}_{zp}(x)+\mathcal{U}_{zp}(L-x)-2\mathcal{U}%
_{zp}(L/2)
\end{equation}
\ Since we are dealing with divergent quantities, we introduce a temporary
high-frequency cut-off; after finding the change in zero-point energy for the
situation including the cut-off, we then take the no-cut-off limit. \ In our
calculation, we also introduce a parameter $s$ so as to treat both the like-
and unlike-boundary conditions at the same time. Specifically, the change in
zero-point energy for the partitioned box with a cut-off parameter $\Lambda$
is\cite{cas}
\begin{align}
\Delta\mathcal{U}_{zp}(x,L,\Lambda)  &  =\sum_{n=1}^{\infty}\frac{1}{2}%
\frac{(n-s)\pi c}{x}\exp\left[  -\Lambda\frac{(n-s)}{x}\right]  +\sum
_{n=1}^{\infty}\frac{1}{2}\frac{(n-s)\pi c}{L-x}\exp\left[  -\Lambda
\frac{(n-s)}{L-x}\right] \nonumber\\
&  -2\sum_{n=1}^{\infty}\frac{1}{2}\frac{(n-s)\pi c}{L/2}\exp\left[
-\Lambda\frac{(n-s)}{L/2}\right] \nonumber\\
&  =-\frac{\pi c}{2}\frac{\partial}{\partial\Lambda}\left\{  \sum
_{n=1}^{\infty}\exp\left[  -\Lambda\frac{(n-s)}{x}\right]  +[x\rightarrow
(L-x)]-2[x\rightarrow L/2]\right\} \nonumber\\
&  =-\frac{\pi c}{2}\frac{\partial}{\partial\Lambda}\left\{  \left[
\exp\left[  \frac{\Lambda s}{x}\right]  \frac{1}{\exp[\Lambda/x]-1}\right]
+[x\rightarrow(L-x)]-2[x\rightarrow L/2]\right\} \nonumber\\
&  =-\frac{\pi c}{2}\frac{\partial}{\partial\Lambda}\left\{  \left[  \frac
{x}{\Lambda}+(s-\frac{1}{2})+\left(  \frac{1}{12}+\frac{s^{2}}{2}-\frac{s}%
{2}\right)  \frac{\Lambda}{x}+...\right]  +[x\rightarrow(L-x),L/2]\right\}
\nonumber\\
&  =-\frac{\pi c}{2}\left(  \frac{1}{12}+\frac{s^{2}}{2}-\frac{s}{2}\right)
\left[  \frac{1}{x}+\frac{1}{L-x}-\frac{4}{L}\right]  +O(\Lambda/x)
\label{DUzp}%
\end{align}
where the parameter $s=0$ or $s=1/2.$ \ In the no-cut-off limit $\Lambda
\rightarrow0,$ the change in zero-point energy becomes
\begin{equation}
\Delta\mathcal{U}_{zp}(x,L)=-\frac{\pi c}{2}\left(  \frac{1}{12}+\frac{s^{2}%
}{2}-\frac{s}{2}\right)  \left[  \frac{1}{x}+\frac{1}{L-x}-\frac{4}{L}\right]
\label{DUzp2}%
\end{equation}
If both the walls and partition require Dirichlet boundary conditions for the
field, (corresponding to the frequency in Eq. (\ref{wDir}) and $s=0$ in Eqs.
(\ref{DUzp}) and (\ref{DUzp2})), then the initial coefficient in Eq.
(\ref{DUzp2}) is negative%
\begin{equation}
\Delta\mathcal{U}_{zp}^{DD}(x,L)=-\frac{\pi c}{24}\left[  \frac{1}{x}+\frac
{1}{L-x}-\frac{4}{L}\right]  \label{UDD}%
\end{equation}
In this case, the partition is attracted to the ends of the large box. \ This
same zero-point energy arises when Neumann boundary conditions are applied for
both the ends of the box and the partition, $\Delta\mathcal{U}_{zp}%
^{DD}(x,L)=\Delta\mathcal{U}_{zp}^{NN}(x,L)$. On the other hand, if the
partition requires Neumann boundary conditions while the ends of the large box
require Dirichlet boundary conditions (corresponding to the frequency in Eq.
(\ref{wNeu}) and $s=1/2$ in Eqs. (\ref{DUzp}) and (\ref{DUzp2}), then the
initial coefficient in Eq. (\ref{DUzp2}) is positive,%
\begin{equation}
\Delta\mathcal{U}_{zp}^{DN}(x,L)=+\frac{\pi c}{48}\left[  \frac{1}{x}+\frac
{1}{L-x}-\frac{4}{L}\right]  \label{UDN}%
\end{equation}
and the partition is repelled by the ends of the large box. \ We can also take
the limit $L\rightarrow\infty$ as the size of the large box becomes infinitely
long. \ Then for the situation of like boundary conditions between the
partition and the wall, we have from Eq. (\ref{UDD})
\begin{equation}
\Delta\mathcal{U}_{zp}^{DD}(x,)=-\frac{\pi c}{24x} \label{DDi}%
\end{equation}
while for unlike boundary conditions between the partition and the wall, we
have from Eq. (\ref{UDN})%
\begin{equation}
\Delta\mathcal{U}_{zp}^{DN}(x,L)=+\frac{\pi c}{48x} \label{DNi}%
\end{equation}

\subsection{Change in Energy for the Rayleigh-Jeans Spectrum}

We can also calculate the change in the radiation energy stored in the box for
the limit of high temperature where the thermal spectrum is expected to
approach energy equipartition $U(\omega,T)\rightarrow k_{B}T.$ \ Since we are
using units where $k_{B}=1,$ this corresponds to $U(\omega,T)\rightarrow T.$
The calculation is carried out in the same style as for zero-point radiation
except that the energy per normal mode is different. \ Thus analogous to Eq.
(\ref{DUzp}), we have%
\begin{align}
\Delta\mathcal{U}_{RJ}(x,L,T,\Lambda)  &  =\sum_{n=1}^{\infty}T\exp\left[
-\Lambda\frac{(n-s)}{x}\right]  +[x\rightarrow(L-x)]-2[x\rightarrow
L/2]\nonumber\\
&  =T\left\{  \left[  \exp\left[  \frac{\Lambda s}{x}\right]  \frac{1}%
{\exp[\Lambda/x]-1}\right]  +[x\rightarrow(L-x)]-2[x\rightarrow L/2]\right\}
\nonumber\\
&  =T\left\{  \left[  \frac{x}{\Lambda}+(s-\frac{1}{2})+\left(  \frac{1}%
{12}+\frac{s^{2}}{2}-\frac{s}{2}\right)  \frac{\Lambda}{x}+...\right]
+[x\rightarrow(L-x),L/2]\right\} \nonumber\\
&  =0+O(\Lambda/x)
\end{align}
In the no-cut-off limit $\Lambda\rightarrow0,~$we find
\begin{equation}
\Delta\mathcal{U}_{RJ}(x,L,T)=0 \label{dRJ}%
\end{equation}
irrespective of the boundary conditions. \ 

We notice that the basic form of the energy changes in these two limiting
situations of zero-point radiation (Eqs. (\ref{UDD})-(\ref{DNi})) and of the
Rayleigh-Jeans spectrum (Eq. (\ref{dRJ})) might be suggested from scaling or
dimensional considerations alone. \ Each piece $\mathcal{U}(x,T),$
$\mathcal{U}(L-x,T),~\mathcal{U}(L/2,T)$ in the energy change $\Delta
\mathcal{U}(x,L,T)$ involves a single length. \ Thus there is no dependence
upon the ratio of lengths $x/L.$ \ The energy must scale as an inverse length.
\ For the Rayleigh-Jeans spectrum where $U_{RJ}(\omega,T)=T,$ the energy must
depend upon the temperature $T$, but there is no connection between length and
temperature. \ Furthermore, when calculating the change of energy
$\Delta\mathcal{U}$, we must introduce a cut-off $\omega_{cut-off}$ in
frequency and make the subtractions before taking the cut-off frequency to
infinity. However, if $\omega_{cut-off}$ is the cut-off frequency, then the
number $n_{x}$ of modes of frequency lower than the cut-off frequency for a
box of length $x$ \ is $n_{x}\pi c/x=\omega_{cut-off},$ while the number of
radiation modes below the cut-off frequency for the part of the box on the
other side of the partition is such that $n_{L-x}\pi c/(L-x)=\omega_{cut-off}%
$. Therefore the total number of modes entering when the partition is located
at position $x$ is
\begin{equation}
n_{x}+n_{L-x}=\frac{x}{\pi c}\omega_{cut-off}+\frac{L-x}{\pi c}\omega
_{cut-off}=\frac{L}{\pi c}\omega_{cut-off}%
\end{equation}
which is independent of the position $x$ of the partition. If each mode makes
the same energy contribution (as is the case for the Rayleigh-Jeans spectrum),
then on subtraction of the energy when the partition is half-way across the
box, the change of energy $\Delta\mathcal{U}_{RJ}$ vanishes. \ In the case of
zero-point radiation, the normal mode energies do indeed depend upon the
frequencies $\omega_{n}$ of the normal modes, and so the energy change is
nonvanishing, and indeed scales as an inverse distance as seen in Eqs.
(\ref{UDD})-(\ref{DNi}). \ 

\subsection{Thermodynamic Minimum Principle for the Helmholtz Free Energy}

The radiation in a partitioned box can be regarded as the working substance
for a thermodynamic system involving length parameters $x$ and $L$ (analogous
to volume) and temperature $T.$\ For a fixed length $L$ and temperature $T,$
the Helmholtz free energy achieves its minimum value at thermal equilibrium.
\ For the case of unlike boundary conditions (where the radiation satisfies
Dirichlet boundary conditions at the ends of the box but Neumann boundaries at
the partition), the situation at zero temperature involves a repulsion of the
partition from the walls so that thermal equilibrium corresponds to the
partition being located in the middle of the box at $x=L/2.$ \ In this
situation of zero-temperature, the Helmholtz free energy $F=U-TS$ equals the
energy since the entropy vanishes at zero temperature. \ From Eq. (\ref{UDN}),
we see that the energy is indeed a minimum for the partition in the middle of
the box. \ For finite non-zero temperature, we expect that the equilibrium
position is still in the center of the box, but the change in the Helmholtz
free energy at other positions will be modified compared to that for zero
temperature. \ In every position $x$ and for any total box length $L$, the
change in the Helmholtz free energy for constant temperature must provide the
pressure which moves the partition invariably toward its thermal equilibrium
position at the center of the box.\cite{Morse} \ This requirement places an
enormous restriction on the functional form of the energy $U_{n}(\omega
_{n},T)$ per normal mode of frequency $\omega_{n}$. \ Taken together with the
asymptotic limits $U(\omega,T)\rightarrow\omega/2$ for $\omega>>T$ and
$U(\omega,T)\rightarrow T$ for $T>>\omega,$ the minimum Helmholtz free energy
condition is sufficient to determine the allowed spectrum of blackbody
radiation. \ We simply assume a test functional form $\phi_{t}(\omega/T)$ for
the canonical potential of a radiation normal mode (which is the same as the
canonical potential for a harmonic oscillator discussed above) which satisfies
the asymptotic limits in Eq. (\ref{mp}). \ This test potential $\phi
_{t}(\omega/T)$ then determines the Helmholtz free energy $F_{tn}(\omega
_{n},T)$\ of each normal mode as $F_{tn}(\omega_{n},T)=-T\phi_{t}(\omega
_{n}/T),$ where the frequency $\omega_{n}$ is related to the length of the box
as in Eq. (\ref{wNeu}) \ Then we proceed to calculate the change in Helmholtz
free energy $\Delta\mathcal{F}_{t}(x,L,T)$ as a function of $x$\ for the
radiation in the partitioned box of length $L$,%
\begin{equation}
\Delta\mathcal{F}_{t}(x,L,T)=\sum_{n=1}^{\infty}F_{tn}(\omega_{n}%
(x))+\sum_{n=1}^{\infty}F_{tn}(\omega_{n}(L-x))-2\sum_{n=1}^{\infty}%
F_{tn}(\omega_{n}(L/2))
\end{equation}
\ Only if the Helmholtz free energy $\Delta\mathcal{F}_{t}(x,L,T)$ so obtained
is a smooth monotonic function which reaches its minimum at $x=L/2$ for all
box lengths $L,$ do we have a possible choice for the spectrum of blackbody
radiation. \ 

In the calculations, it seems easiest to first separate off the divergent
zero-point energy, calculate the change in the remaining convergent series,
and then add back the change in Helmholtz free energy which is associated with
the zero-point energy. \ Numerical calculation easily shows that the Planck
formula including zero-point radiation given in Eq. (\ref{phi}) leads to a
change in Helmholtz free energy at fixed temperature which meets all the
required conditions. \ In Fig. 1, we give a graph of the change in the
Helmholtz free energy $\Delta\mathcal{F}_{Pzp}(x,L,T)$ for a box of total
length $L=5,$ and temperatures $T=0,~1,~3.$ \ The Planck formula indeed
satisfies the \textquotedblleft minimum\textquotedblright\ behavior for the
Helmholtz free energy which is required by thermodynamics.

On the other hand, for all the test functions $\phi_{t}(\omega/T)$ which met
the asymptotic conditions on the canonical potential but which departed from
the Planck formula, it was easy to show that the change in Helmholtz free
energy associated with these functions did not provide a monotonic function of
$x$ for some total box length $L.$ \ In Fig. 2, we give a graph of the test
canonical function $\phi_{t}(\omega/T)=\{-\ln[2\sinh(\omega/2T)]+\omega
/(2T)\}\exp[-\omega/T]-\omega/(2T)$ which meets the asymptotic conditions
(\ref{mp}) for the thermodynamics of the harmonic oscillator but does not
match the Planck function, and clearly does not meet the thermodynamic
requirements for the change in the Helmholtz free energy for the partitioned
box. \ 

It is our conclusion, that thermodynamic arguments provide a basis for the
derivation of the blackbody radiation spectrum.

\section{Zero-Point Energy is Embedded in the Traditional Planck Spectrum}

\subsection{The Traditional Planck Spectrum Omits Zero-Point Radiation}

All textbooks of modern physics and most physicists present the Planck
spectrum without including the zero-point radiation part.\cite{texts} \ Thus
the Planck energy for a harmonic oscillator is usually given as%
\begin{equation}
U_{P}(\omega,T)=\frac{\hbar\omega}{\exp[\hbar\omega/(k_{B}T)]-1}=\frac
{\hbar\omega}{2}\coth\left(  \frac{\hbar\omega}{2k_{B}T}\right)  -\frac
{\hbar\omega}{2} \label{tP}%
\end{equation}
Planck's determination of the blackbody spectrum followed the experimental
work of Lummer and Pringsheim\cite{LP} which measured the random radiation of
a source which was above the random radiation surrounding the detector; the
zero-point radiation which surrounded a source also surrounded the detector
and so was not measured. It is only recently that we have experimental
measurements of Casimir forces\cite{Cf} which measure all the radiation
surrounding the parallel plates. \ 

Because the traditional Planck formula in (\ref{tP}) omits the zero-point
radiation, discussions of the blackbody radiation usually make no reference to
zero-point radiation. \ On the other hand, all the derivations of the
blackbody radiation spectrum within classical physics depend crucially upon
the presence of zero-point radiation. \ In the present article, we have given
the basis for two derivations of the blackbody spectrum making use of
thermodynamic ideas, and the existence of zero-point radiation in the
low-temperature asymptotic limit is crucial to the discussions. \ 

\subsection{The Change in Radiation Energy in a Partitioned Box Reveals the
Zero-Point Energy Hidden in the Traditional Planck Spectrum}

According to the traditional Planck formula in Eq. (\ref{tP}), the
high-temperature limit for the energy of a harmonic oscillator of frequency
$\omega$ does not go over fully to the equipartition value $k_{B}T$, but
rather retains a finite correction $(1/2)\hbar\omega$ associated with the
absence of the zero-point energy contribution. \ Thus for $k_{B}T>>\hbar
\omega,$ we have from Eq. (\ref{tP})%
\begin{equation}
U_{P}(\omega,T)\rightarrow\hbar\omega\left[  \frac{k_{B}T}{\hbar\omega}%
-\frac{1}{2}+\frac{1}{12}\left(  \frac{k_{B}T}{\hbar\omega}\right)
^{2}-...\right]  =k_{B}T-\frac{1}{2}\hbar\omega+\omega O(\omega/T) \label{Tmz}%
\end{equation}
This failure of the equipartition limit does not seem to bother physicists.
However, the failure of this limit becomes glaringly obvious if we calculate
the change in Casimir energy $\Delta\mathcal{U}_{P}$ associated with the use
of the traditional Planck formula; the large equipartition $k_{B}T$ in Eq.
(\ref{Tmz}) makes no contribution to $\Delta\mathcal{U}_{P}$ leaving only the
negative zero-point result.

The Casimir energy change takes its simplest form when the total length $L$ of
the box goes to infinity, $L\rightarrow\infty$. \ This is seen for the
zero-temperature case in the transition from Eqs. (\ref{UDD}) and (\ref{UDN})
over to Eqs. (\ref{DDi}) and (\ref{DNi}). \ We will take the case where
Dirichlet boundary conditions are applied at both the walls and the partition
so that the frequencies of the radiation normal modes are given in
(\ref{wDir}). \ \ 

The thermal energy $\mathcal{U}_{P}(x,T)$ follows from the Euler-Maclaurin
summation formula,\cite{E-M}%
\begin{equation}
\sum_{k=1}^{n-1}f_{k}=\int_{0}^{n}f(k)dk-\frac{1}{2}[f(0)+f(n)]+\frac{1}%
{12}[f^{\prime}(n)-f^{\prime}(0)]-\frac{1}{720}[f^{\prime\prime\prime
}(n)-f^{\prime\prime\prime}(0)]+R \label{E-M}%
\end{equation}
where $R$ is a remainder term and the coefficient terms involve the same
Bernoulli numbers $B_{n}$ as appear in a power-series expansion of the Planck
formula.\cite{Bern} \ For the traditional Planck formula in Eq. (\ref{tP}), we
have for the thermal energy $\mathcal{U}_{P}(x,T)$ in the box of length
$x$\cite{int}%

\begin{align}
\mathcal{U}_{P}(x,T)  &  =\sum_{n=1}^{\infty}\frac{\hbar\omega_{n}}{\exp
[\hbar\omega_{n}/(k_{B}T)]-1}=\sum_{n=1}^{\infty}\frac{\hbar(n\pi c/x)}%
{\exp[\hbar n\pi c/(k_{B}Tx)]-1}\nonumber\\
&  =\int_{0}^{\infty}dn\frac{\hbar(n\pi c/x)}{\exp[\hbar n\pi c/(k_{B}%
Tx)]-1}-\frac{k_{B}T}{2}+\frac{\pi c}{24x}+TO(1/T^{2}x^{2})\nonumber\\
&  =\frac{\pi}{6}\frac{(k_{B}T)^{2}}{\hbar c}x-\frac{k_{B}T}{2}+\frac{\pi
c}{24x}+TO(1/T^{2}x^{2}) \label{Ux}%
\end{align}
Thus the change in the thermal energy $\Delta\mathcal{U}_{P}(x,T)$ associated
with the partition position $x$ in the limit $L\rightarrow\infty$ becomes
\begin{align}
\Delta\mathcal{U}_{P}(x,T)  &  =\lim_{L\rightarrow\infty}\Delta\mathcal{U}%
_{T}(x,L,T)=\lim_{L\rightarrow\infty}[\mathcal{U}_{T}(x,T)+\mathcal{U}%
_{T}(L-x,T)-2\mathcal{U}_{T}(L/2T)]\nonumber\\
&  =\mathcal{U}_{T}(x,T)+\lim_{L\rightarrow\infty}\left[  \frac{\pi}{6}%
\frac{(k_{B}T)^{2}}{\hbar c}(L-x)-\frac{k_{B}T}{2}+\frac{\pi c}{24(L-x)}%
-...\right] \nonumber\\
&  -\lim_{L\rightarrow\infty}2\left[  \frac{\pi}{6}\frac{(k_{B}T)^{2}}{\hbar
c}\frac{L}{2}-\frac{k_{B}T}{2}+\frac{\pi c}{24(L/2)}-...\right] \nonumber\\
&  =\mathcal{U}_{P}(x,T)-\left(  \frac{\pi}{6}\frac{x(k_{B}T)^{2}}{\hbar
c}x-\frac{k_{B}T}{2}\right)  \label{DUx}%
\end{align}
Now introducing the Euler-Maclaurin expansion for a finite-length box given in
Eq. (\ref{Ux}) into Eq. (\ref{DUx}), we have
\begin{equation}
\Delta\mathcal{U}_{P}(x,T)=\frac{\pi c}{24x} \label{DUT}%
\end{equation}
provided that the the remainder $R$ in the Euler-Maclaurin summation formula
is small. \ But $\pi c/(24x)$ is exactly the negative of the change of
zero-point energy in Eq. (\ref{DDi}) for the box of length $x.$ \ Thus the
traditional Planck expression in Eq. (\ref{tP}) which has no zero-point energy
at low temperature betrays its connection to zero-point energy by giving at
high temperatures a change in energy $\Delta\mathcal{U}_{P}(x,T)$ which is the
negative of the change in zero-point energy.

The evaluation in Eq. (\ref{DUT}) giving the connection to zero-point energy
holds for situations $xT>>1$ where the remainder term $R$ for Euler-Maclaurin
summation formula has only a small value in Eq. (\ref{Ux}). \ For values of
$xT$ $\lesssim1,$ the remainder $R$\ becomes relatively large and the
Euler-Maclaurin summation formula in (\ref{E-M}) does not give a good
approximation to the change in the thermal radiation energy in the region of
length $x.$ \ 

Figure 3 shows the changes in Casimir energy $\Delta\mathcal{U}_{P}^{DD}(x,T)$
when Dirichlet boundary conditions are applied at both the end of the box and
at the partition, for three different functions in the limit of an infinitely
long box $L\rightarrow\infty$. \ The three functions involve \ a) the Planck
spectrum including zero-point energy $\Delta\mathcal{U}_{Pzp}(x,T)$, b) the
traditional Planck spectrum without zero-point energy $\Delta\mathcal{U}%
_{P}(x,T)$, and c) zero-point energy $\Delta\mathcal{U}_{zp}(x)$. \ It is
clear that the energy change for the traditional Planck formula which omits
zero-point energy goes over to the negative of the change of zero-point energy
at large values of $xT.$ \ On the other hand, the Planck formula which
includes zero-point energy approaches zero extremely rapidly at large values
of $xT.$ \ This vanishing change of energy is consistent with the
Rayleigh-Jeans spectrum as the high-temperature limit of the blackbody
spectrum where $\Delta\mathcal{U}_{RJ}(x,T)=0$. \ In the high-temperature
Casimir energy changes, we see clear evidence that the idea of zero-point
energy is embedded in the traditional Planck formula despite the explicit
removal of the zero-point energy from the low-temperature limit.

\section{Discussion: Connections Between Thermal Radiation,
Zero-Point-Radiation, and Spacetime Structure}

The Planck spectrum of thermal radiation within classical physics is
intimately connected with the spectrum of classical zero-point radiation and
with the structure of spacetime. This striking idea seems rarely appreciated
among physicists today. In the past, this connection has been derived using
non-inertial coordinate frames. Specifically, it has been shown that the
correlation function for the zero-point radiation fields depends only upon the
geodesic separation between the spacetime points where the correlation
function is evaluated.\cite{st} In Minkowski spacetime,\cite{Min} for example,
the field correlation functions depend upon the Lorentz-invariant spacetime
interval $(ct-ct^{\prime})^{2}-(\mathbf{r}-\mathbf{r}^{\prime})^{2}.$
\ Furthermore, thermal radiation can be derived from zero-point radiation by
the use of a time-dilating conformal transformation in a non-inertial
frame.\cite{acc}

In the present work, we point out that the use of thermodynamic ideas in
connection with classical zero-point radiation leads naturally to the Planck
spectrum for blackbody radiation, for both an individual radiation mode and
for the Casimir energy change for radiation in a partitioned box. \ The Planck
spectrum appears if one requires that the interpolation between the zero-point
energy at low temperature and the equipartition energy at high temperature is
thermodynamically smooth in the sense that the canonical potential function
$\phi(\omega/T)$ (which for an oscillator depends upon one variable) is
monotonic and all its derivatives are monotonic so that no preferred value of
$\omega/T$ is singled out. \ Also, the Planck spectrum appears if one uses
these same asymptotic limits for a single radiation mode but requires that at
fixed temperature the Helmholtz free energy in a partitioned box assumes its
minimum value at thermal equilibrium. \ In addition, the zero-point energy is
embedded even in the traditional Planck spectrum which omits zero-point energy
as the low temperature limit; the zero-point energy reappears in the
high-temperature limit for both a single oscillator (or radiation mode) and
also for the change in Casimir energy for radiation in a partitioned box.
\ Indeed, thermal radiation, zero-point radiation, and spacetime structure are
all related.

\section{Acknowledgements}

I wish to thank Professors Nicholas Giovambattista, V. Paramewaran Nair, and
Joel Gersten for helpful discussions.

\bigskip

Figure Captions

\bigskip

Fig. 1. \ Change in Helmholtz Free Energy with Partition Position at Constant Temperature

The change in Helmholtz free energy $\Delta\mathcal{F}_{Pzp}(x,L,T)$ is
plotted as a function of partition position $x$ at three different
temperatures for $0<x\leq2.5$ in a box of length $L=5$ (for unlike boundary
conditions). \ The Helmholtz free energy change $\Delta\mathcal{F}%
_{Pzp}(x,L,T)$ is obtained from the Planck expression (with zero-point energy)
for the canonical potential $\phi_{Pzp}(\omega/T)=-\ln[2\sinh(\omega/2T)]$
given in Eq. (\ref{phi}) for each radiation mode of frequency $\omega.$ a) The
solid curve corresponds to temperature $T=0$. \ b) The dashed curve is for
$T=1.$ \ c) The dotted curve is for $T=3.$ \ The curves show a monotonic
decease toward 0 at the middle of the box $x=2.5,$ consistent with
thermodynamic requirements.

\bigskip

Fig. 2. \ Change in Helmholtz Free Energy Assuming a Test Radiation Spectrum
Different from the Planck Spectrum

The change in Helmholtz free energy $\Delta\mathcal{F}_{t}(x,L,T)$ is plotted
as a function of partition position $x$ when we consider a test spectrum
different from the Planck spectrum (for unlike boundary conditions). \ Here
the canonical potential for each radiation mode of frequency $\omega$ at
temperature $T>0$ is chosen as $\phi_{t}(\omega/T)=\{-\ln[2\sinh
(\omega/2T)]+\omega/(2T)\}\exp(-\omega/T)-\omega/(2T).$ \ The box has total
length $L=5,$ and the plot shows half the box, $0<x\leq2.5.$ \ a) The solid
curve curve corresponds to the energy change for zero-point energy at zero
temperature$.$ \ b) The dashed curve is for $T=1$. \ c) The dotted curve is
for $T=3.$ \ The curves for temperature $T>0$ do not show monotonic behavior
and so violate the thermodynamic requirements. \ Therefore the assumed test
canonical potential $\phi_{t}$\ cannot correspond to thermal radiation.

\bigskip

Fig. 3. \ Change in Energy with Partition Position at Constant Temperature

The change in energy $\Delta\mathcal{U}(x,T)$ is plotted as a function of
partition position $x$ for constant temperature for an infinitely long box,
$L\rightarrow\infty$ (for like boundary conditions). \ \ a) The solid curve
gives the energy change $\Delta\mathcal{U}_{Pzp}(x,T)=\Delta\mathcal{U}%
_{P}(x,T)+\Delta\mathcal{U}_{zp}(x)$ at $T=1$ following from the full Planck
spectrum (which includes zero-point radiation) given in Eq. (\ref{UPzp}).
\ \ b) The dashed curve gives the thermal energy change $\Delta\mathcal{U}%
_{P}(x,T)$ at $T=1$ following from the traditional Planck spectrum in Eq.
(\ref{tP}) (which omits zero-point energy) for each mode. c) The dotted curve
gives the zero-point energy change $\Delta\mathcal{U}_{zp}(x)$ in Eq.
(\ref{DDi}). \ For $xT>>1,$ the traditional Planck spectrum without zero-point
energy gives a change in energy $\Delta\mathcal{U}_{P}$ which is the negative
of the zero-point energy change $\Delta\mathcal{U}_{zp}$. \ For $xT>>1,$ only
the full Planck spectrum with zero-point energy goes over to the expected
energy change $\Delta\mathcal{U}_{RJ}(x,T)=0$ holding for the Rayleigh-Jeans spectrum.


\begin{thebibliography}{99}
\bibitem {Bern}The Planck expression without zero-point energy in Eq.
(\ref{tP}) can be expanded as $t/(e^{t}-1)=\sum_{n=0}^{\infty}B_{n}%
t^{n}/n!=1-t/2+t^{2}/12-t^{4}/720+...$ for $|t|<2\pi.$ \ See Abramowitz and
Stegun in ref. 17, p. 804, \#23.1.1.
\end{thebibliography}
\end{document}